%% file: main.tex
\documentclass{article} 
\usepackage{iclr2026_conference,times}

\input{math_commands.tex}

\usepackage{hyperref}
\usepackage{url}

\usepackage{subcaption}
\usepackage{graphicx}
\usepackage{amsmath}
\usepackage{siunitx}
\usepackage{algorithm}
\usepackage{algpseudocode}  
\usepackage{booktabs}
\usepackage{multirow}
\usepackage{array}
\usepackage{float}  
\usepackage{tabularx}

\usepackage[section]{placeins}

\newcolumntype{Y}{>{\centering\arraybackslash}X}

\title{AB-PIELMs: Adaptive-Basis Physics-Informed Extreme Learning Machines for Residual-Driven Domain Decomposition}
\author{Aldis Daniel\\
Department of Mechanical Engineering\\
Indian Institute of Technology Madras\\
Chennai 600036, India \\
\texttt{me22b070@smail.iitm.ac.in}
\AND
Vikas Dwivedi\\
CREATIS, INSA-Lyon; Inserm, U1044; CNRS UMR 5220\\
Université Lyon 1; Université de Lyon\\
Lyon, 69621, France\\
\texttt{vikas.dwivedi@creatis.insa-lyon.fr}
\AND
Balaji Srinivasan\\
Wadhwani School of Data Science and Artificial Intelligence\\
Indian Institute of Technology Madras\\
Chennai 600036, India \\
\texttt{sbalaji@dsai.iitm.ac.in}
}

\iclrfinalcopy 
\begin{document}

\maketitle

\begin{abstract}
Physics-informed neural solvers often struggle with multiscale behavior and sharp gradients due to the use of global approximation spaces and nonconvex training. We introduce AB-PIELM, an adaptive-basis physics-informed extreme learning machine that combines domain decomposition with joint optimisation of spatial and spectral hyperparameters while retaining deterministic normal-equation training. The method adapts both the number and placement of subdomains and the locality of radial basis functions, enabling problem-dependent allocation of representational capacity without modifying the underlying solver. Experiments on oscillatory function approximation and singularly perturbed advection–diffusion equations demonstrate accurate solutions across a wide range of stiffness regimes while using substantially fewer neurons than existing neural and PIELM-based approaches. The learned hyperparameters are interpretable, concentrating resolution near sharp gradients. The framework also supports inverse problems, successfully recovering diffusion coefficients from sparse noisy observations using Bayesian optimisation. These results indicate that adaptive-basis PIELM formulations provide an efficient and stable alternative to gradient-trained neural PDE solvers and naturally extend to broader classes of differential equations.
\end{abstract}

\section{Introduction}
\label{sec:intro}
Recent years have seen rapid growth in machine-learning approaches for solving partial differential equations (PDEs) 
and surrogate modeling of physical systems. Among these, physics-informed neural networks (PINNs) embed governing 
equations and boundary conditions directly into the loss function, enabling data-efficient learning without labeled 
solutions \citep{raissi_physics-informed_2019}. Subsequent developments have applied physics-informed learning to 
forward and inverse problems, parametric studies, and uncertainty quantification 
\citep{karniadakis_physics-informed_2021}. However, most neural PDE solvers rely on global function approximators 
defined over the entire spatial domain. Such representations implicitly assume uniform smoothness and are 
therefore sensitive to stiffness, multiscale structure, and sharp gradients. As a result, PINNs may exhibit slow 
convergence or failure to train, caused by gradient pathologies and ill-conditioned optimization landscapes 
\citep{wang_understanding_2021, krishnapriyan_characterizing_2021, wang_when_2022}. From a spectral perspective, 
this behavior reflects the low-frequency bias of neural networks, which limits the efficient representation of 
localized high-frequency features \citep{rahaman_spectral_2019}.

To overcome these limitations, several works introduce localized neural approximations. Domain-decomposition 
methods such as cPINNs, XPINNs, and related parallel formulations train multiple networks on subdomains coupled 
through interface conditions \citep{jagtap_conservative_2020, d_jagtap_extended_2020, shukla_parallel_2021, 
klawonn_machine_2024}. Other approaches incorporate 
spectral enrichment, for example via Fourier feature mappings \citep{tancik_fourier_2020}, or adaptive 
decomposition strategies \citep{botvinick-greenhouse_ab-pinns_2025}. In parallel, deterministic alternatives 
replace gradient training with linear-in-parameter models. Extreme Learning Machines (ELMs) fix nonlinear feature 
maps and solve for coefficients directly \citep{huang_extreme_2006}, and physics-informed extensions such as the 
normal-equation PIELM (NEPIELM) provide stable solvers that admit domain-decomposed formulations 
\citep{dwivedi_normal_2022,anderson_elm-fbpinn_2024,dwivedi_gated_2025}. While these methods improve robustness 
relative to global networks, domain partitions and basis parameters are typically prescribed, limiting adaptation 
to localized solution features.

Despite these advances, existing approaches either rely on nonconvex optimization or require manually designed 
decompositions and hyperparameters, leading to complicated training procedures, increased computational cost, 
and reduced interpretability compared to traditional numerical discretizations. In this work we propose AB-PIELM, a 
parameterisable domain-decomposed physics-informed extreme learning machine that explicitly identifies spatial 
and spectral hyperparameters and optimizes them within a single physics-informed objective, while retaining 
deterministic normal-equation training for fixed hyperparameters. The computational efficiency of the underlying 
PIELM allows for optimisation of the network architecture whereas existing methods rely on residual-based refinement 
of collocation distributions.

The main contributions of this work are:
\begin{itemize}
\item We identify key spatial and spectral hyperparameters in the formulation and propose their joint optimization within a single physics-informed objective.
\item We rigorously compare our method against state of the art approaches like XTFC and distributed NE-PIELM on a benchmark ODE.
\item We demonstrate the interpretability of the resulting domain partitions and basis parameters.
\item We formulate the inverse problem and solve it using Bayesian optimisation, demonstrating the method's flexibility.
\end{itemize}

The results shown are in benchmark cases as a proof of concept and the approach naturally extends to PDEs. The code is 
made available at \url{https://github.com/catdisk04/AB-PIELM} to facilitate future research. 


\section{Methodology}
\label{sec:methodology}

The proposed method combines the ideas of Adaptive Mesh Refinement (AMR) and modern hyperparamter optimisation techniques
in the context of PIELMs. This allows for a more flexible and adaptive architecture that can better capture sharp gradients 
and complex solution features while retaining the computational efficiency of PIELMs that are trained using the normal equation 
formulation \citep{dwivedi_normal_2022}. 

\subsubsection{AB-PIELM Solution Scheme}
\label{methodology:ab_pielm_solution_scheme}

In this section, we provide a brief overview of the distributed formulation used in this work.

We consider linear partial differential equations of the form
\begin{equation}
\mathcal{L}u(x) = f(x), \qquad x \in \Omega \subset \mathbb{R}^d,
\end{equation}
subject to boundary conditions
$
\mathcal{B}u(x) = g(x)$ over $ x \in \partial\Omega,
$
where $\mathcal{L}$ is a linear differential operator and $\mathcal{B}$ is a
boundary operator. The domain $\Omega$ is partitioned into $M$ non-overlapping 
subdomains $\{\Omega_m\}_{m=1}^M$ with interfaces $\Gamma_{mn} = \partial\Omega_m \cap \partial\Omega_n$. 
Each network has output layer weights $\beta$ and hidden layer basis functions $\Phi$ that are local to the subdomain. The solution in each subdomain is given by

\begin{equation}
\hat{u}_m(x) = \Phi_m(x)\beta_m, \qquad x \in \Omega_m.
\end{equation}

Let $\{x_i^{(m)}\}$ denote collocation points in $\Omega_m$.
We define the discrete PDE operator $P_m$ and the discrete boundary operator $B_m$ by
\begin{equation}
P_m \beta_m =
\big(\mathcal{L}\Phi_m(x_i^{(m)})\beta_m\big)_{i},
\end{equation}
\begin{equation}
B_m \beta_m =
\big(\mathcal{B}\Phi_m(x_i^{(m)})\beta_m\big)_{i}.
\end{equation}

For each interior interface $\Gamma_{mn}$, we define trace and flux operators
\begin{align}
C_{m|n}\beta_m &= \Phi_m(x)\beta_m \big|_{x \in \Gamma_{mn}}, \\
D_{m|n}\beta_m &= \partial_n \Phi_m(x)\beta_m \big|_{x \in \Gamma_{mn}},
\end{align}
where $\partial_n$ denotes the outward normal derivative on $\partial\Omega_m$.

The global objective function is defined as 
\begin{align}
J(\beta)
&=
J_{PDE}(\beta) + J_{BC}(\beta) + J_{INT}(\beta),
\label{eq:global_objective}
\end{align}
where

\begin{align}
J_{PDE}(\beta) &= \sum_{m=1}^M \|P_m\beta_m - f_m\|_2^2, \\
J_{BC}(\beta) &= \gamma_b \sum_{m=1}^M \|B_m\beta_m - g_m\|_2^2, \\
J_{INT}(\beta) &= \sum_{(m,n)\in\mathcal{I}} \Big(
\gamma_c \|C_{m|n}\beta_m - C_{n|m}\beta_n\|_2^2
+ \gamma_d \|D_{m|n}\beta_m - D_{n|m}\beta_n\|_2^2
\Big),
\end{align}

and $\mathcal{I}$ denotes the set of neighboring subdomain pairs and
$\gamma_b,\gamma_c,\gamma_d > 0$ are penalty parameters.

The first-order optimality condition $\nabla_\beta J = 0$ yields the block linear
system
\begin{equation}
A\beta = b,
\end{equation}
with diagonal blocks
\begin{equation}
A_{mm}
=
P_m^\top P_m
+ \gamma_b B_m^\top B_m
+ \sum_{n\in\mathcal{N}(m)}
\left(
\gamma_c C_{m|n}^\top C_{m|n}
+ \gamma_d D_{m|n}^\top D_{m|n}
\right),
\end{equation}
and off-diagonal coupling blocks
\begin{equation}
A_{mn}
=
-
\left(
\gamma_c C_{m|n}^\top C_{n|m}
+ \gamma_d D_{m|n}^\top D_{n|m}
\right),
\qquad m \neq n .
\end{equation}
The right-hand side is given by
\begin{equation}
b_m = P_m^\top f_m + \gamma_b B_m^\top g_m .
\end{equation}

The resulting system is symmetric and positive semi-definite, with block sparsity
induced by the domain decomposition. The system is solved directly as 

\begin{equation}
\beta = A^\dagger b,
\end{equation}
where $A^\dagger$ denotes the Moore-Penrose pseudoinverse of $A$.

\subsection{Joint spatial--spectral optimization}

The PDE loss (Eq \ref{eq:global_objective}) of the PIELM network depends on the set of structural
hyperparameters $\theta = (\theta_{\mathrm{spatial}},\theta_{\mathrm{spectral}})$.
The spatial hyperparameters specify the domain decomposition via the interface
locations
\begin{equation}
\theta_{\mathrm{spatial}} = \{x_{\mathrm{if}}^{(k)}\}_{k=1}^{M-1},
\end{equation}
where the number of domains $M$ is also a hyperparameter, thereby defining the subdomains $\Omega_m(\theta_{\mathrm{spatial}})$ . The spectral hyperparameters control the
locality of the RBF basis in each subdomain through the width parameters
\begin{equation}
\theta_{\mathrm{spectral}} = \{\sigma_m\}_{m=1}^M.
\end{equation}
\begin{equation}
\sigma_m = \alpha_m \, \Delta c_m ,
\end{equation}
where $\Delta c_m$ denotes RBF center spacing and $\sigma_m$ the RBF widthin $\Omega_m$.

The loss term $J_{PDE}(\beta;\theta)$ therefore depends on $\theta$ through
both the discrete interface operators and the basis functions of the PIELM formulation. We jointly optimize
\begin{equation}
\min_{\theta \in \Theta} \; \min_{\beta} \; J(\beta;\theta),
\end{equation}
where $\Theta$ is the hyperparameter search space. For fixed $\theta$, the inner minimization is solved exactly 
as described for the underlying PIELM. This joint optimization adapts both the domain
decomposition and the spectral sharpness of the basis in a problem-dependent
manner, concentrating resolution near localized features while preserving
global coherence across interfaces. The dimensionality of $\beta$, $\theta_{\mathrm{spatial}}$, 
and $\theta_{\mathrm{spectral}}$ are dependant on $M$. Hence, optimisation of $M$ is separated from the optimisation
of $\theta$ and executed in an outer loop in an incremental fashion to result in the optimal number of domains $M^*$ 
being as small as possible in the interest of model simplicity.

Since each local network has a fixed number of RBFs and collocation points, 
this framework results in adaptiveness in three levels: The number of domains, 
number of RBFs and the total number of collocation points. The methodology is summarised in Algorithm \ref{alg:dvarpielm}.

\begin{algorithm}[H]
\caption{AB-PIELM Hyperparameter Optimisation Scheme}
\begin{algorithmic}[1]
  \Require Boundary Conditions
  \Require Hyperparameter search space
  \Require Initialise $M$, $J_{min}$, $J_{min}^M$ and $\theta$
  \While{not \textit{converged}} 
    \While{$N_{iter} < N_{max}$} 
      \State Generate collocation set
      \State $\beta^* = \argmin_{\beta} J_{PDE}(\beta;\theta)$ (Solve for output layer weights using normal equation)
      \State Obtain $J$ from the trained optimal network
      \State Evaluate performance on testing set
      \State Update hyperparameters using TPE algorithm
      \If {$J < J_{min}$}
        \State Update $J_{min}$ = $J$ 
        \State Store optimal hyperparameters $\theta^*$
      \EndIf
    \EndWhile
    \If {$J_{min} < J_{min}^M$}
      \State Update $J_{min}^M$ and store optimal hyperparameters $\theta^*$ for $M$
      \State Increment $M$ 
      \State Reset $J_{min}$
    \Else
      \State Update \textit{converged} to True
      \State Store optimal hyperparameters $\theta^*$ for $M^*$
      \State \textbf{break}
    \EndIf
  \EndWhile
  \State Obtain final trained model with optimal hyperparameters $\theta^*$ and $M^*$
  \State Evaluate performance on validation set
  \State \Return Trained AB-PIELM model
\end{algorithmic}
\label{alg:dvarpielm}
\end{algorithm}  

\section{Numerical Experiments}
\label{sec:experiments}

To evaluate the effectiveness and robustness of our approach, we consider two cases: 
\begin{enumerate}
\item Oscillatory function \citep{taylor_optimizing_2022}
\item Singularly perturbed differential equation \citep{de_florio_physics-informed_2024}
\end{enumerate}
The \emph{sinc} function and the advection–diffusion equation serve as complementary benchmark problems. The oscillatory \emph{sinc} 
function with decaying amplitude evaluates the expressive capacity and training behaviour of the approximation architecture, while the 
advection–diffusion equation — a canonical singularly perturbed model — assesses the ability of the domain-decomposed PIELM to 
approximate PDE solutions under stiffness and varying Péclet numbers while maintaining interface continuity and differentiability across 
subdomains.

\subsection{Problem 1: Sinc Function Approximation}
\label{subsec:sinc}

This test case considers the scalar \emph{sinc} function for a function approximation benchmark. In this work, the normalized \emph{sinc}
 function is defined as
\begin{equation}
  sinc(x) =
  \begin{cases}
    1, & x = 0, \\
    \dfrac{\sin(n \pi x)}{n \pi x}, & x \neq 0,
  \end{cases}
  \label{eq:sinc_def}
\end{equation}

where $n = 10$, $x \in [-1.5, 1.5]$

The loss is purely data-driven and constructed from sampling 
$sinc(x)$ at a set of collocation points. This setup isolates the effect of the architecture and domain decomposition on 
approximating an oscillatory function and thereby provides an useful analsis of how the number of neurons and the 
interface location influence the representation of oscillatory data.


\subsection{Problem 2: Advection–Diffusion}
 \label{subsec:adv_diff_sharp}

 The one-dimensional advection diffusion equation is defined as follows: 

\begin{equation}
    - \nu \, \frac{\mathrm{d}^2 u}{\mathrm{d}x^2}
    +\frac{\mathrm{d}u}{\mathrm{d}x}
    = 0,
    \qquad x \in (0,1),
    \label{eq:adv_diff_steady}
  \end{equation}
  with the Dirichlet boundary conditions
  \begin{equation}
    u(0) = 0,
    \qquad
    u(1) = 1,
    \label{eq:adv_diff_bc}
  \end{equation}
  where $\nu > 0$ is the diffusion coefficient.


 
\subsection{Inverse Problem Setup}

We aim to identify the diffusion coefficient $\nu$ in the advection-diffusion equation \eqref{eq:adv_diff_steady} by fitting the 
model to synthetic data that is sparse and noisy, generated from a known value of $\nu$. 

For a fixed $\nu$ and model hyperparamters determined as defined for the forward problem, the trained output layer weights $\beta$ 
minimise the physics-informed loss and results in 

\begin{equation}
    \hat{\textbf{u}}(\nu) = \Phi_{obs}(\nu) \beta(\nu)
\end{equation}

where $\Phi_{obs}(\nu)$ is the matrix of basis functions evaluated at the observation points. The inverse problem is then formulated as the
minimisation of a outer objective

\begin{equation}
    J_{inv}(\nu) = \| \hat{\textbf{u}}(\nu) - \textbf{u}_{data} \|_2^2 + \lambda_{PDE} J_{PDE}(\beta(\nu);\nu),
\end{equation}

where the $J_{PDE}$ part of the objective contributed by the discrete PDE operator and $\lambda_{PDE}$ 
is a penalty parameter that balances the data fit and physics-informed terms. The inverse problem is 

\begin{equation}
    \nu^* = \argmin_{\nu \in {(\nu_{min}, \nu_{max})}} J_{inv}(\nu)
\end{equation}

where $\nu_{min}$ and $\nu_{max}$ define the user-defined search space for the physical parameter $\nu$. The outer optimization is performed using Bayesian optimization with a TPE algorithm. 
This nests the forward model in a outer inverse optimisation loop, allowing for parameter recovery that is simultaneously 
constrained by data and physics.

\section{Results and Discussion}

\subsection{Problem 1: \emph{sinc} Function Approximation}

\begin{figure}[h]
  \centering
  \includegraphics[width=0.5\linewidth]{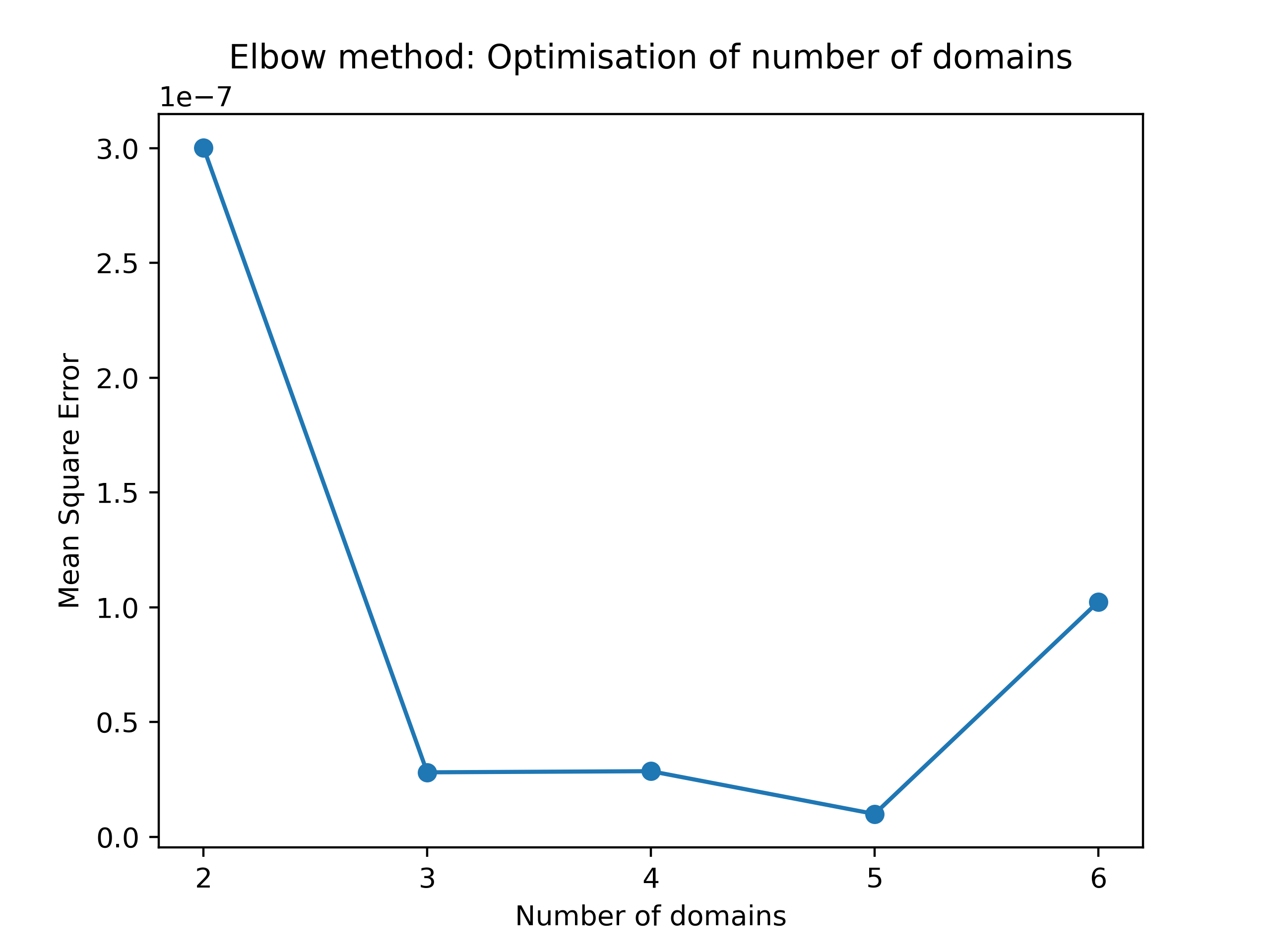}%
  \caption{Plot showing the optimisation the number of subdomains on the MSE residual for the \emph{sinc} function approximation using the Elbow method.}
  \label{fig:sinc_elbow}
\end{figure}
\begin{figure}[h]
  \begin{subfigure}{0.49\linewidth}
    \centering
    \includegraphics[width=\linewidth]{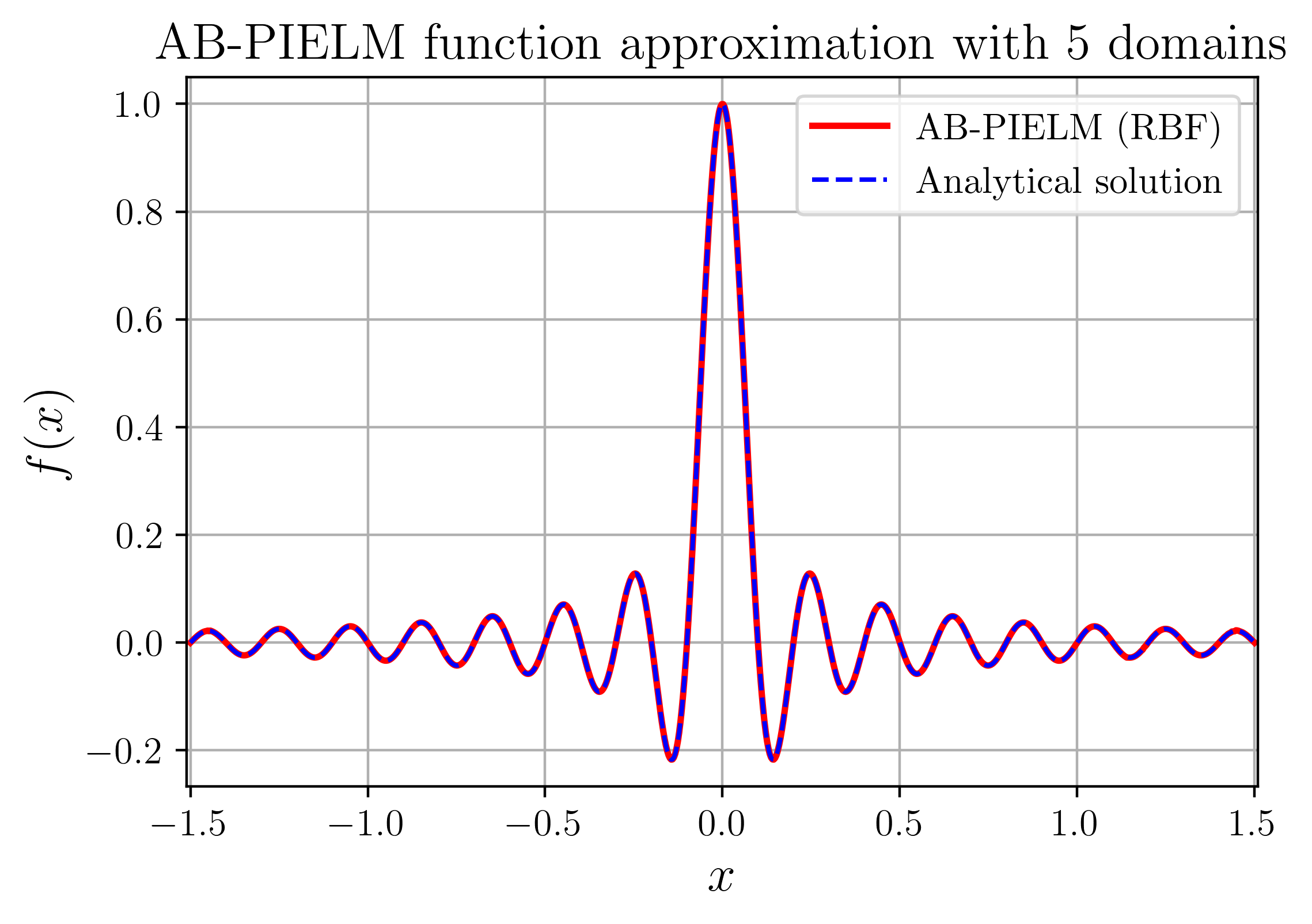}%
    \caption{}
    \label{fig:sinc_5domain}
  \end{subfigure}
  \begin{subfigure}{0.49\linewidth}
    \centering
    \includegraphics[width=\linewidth]{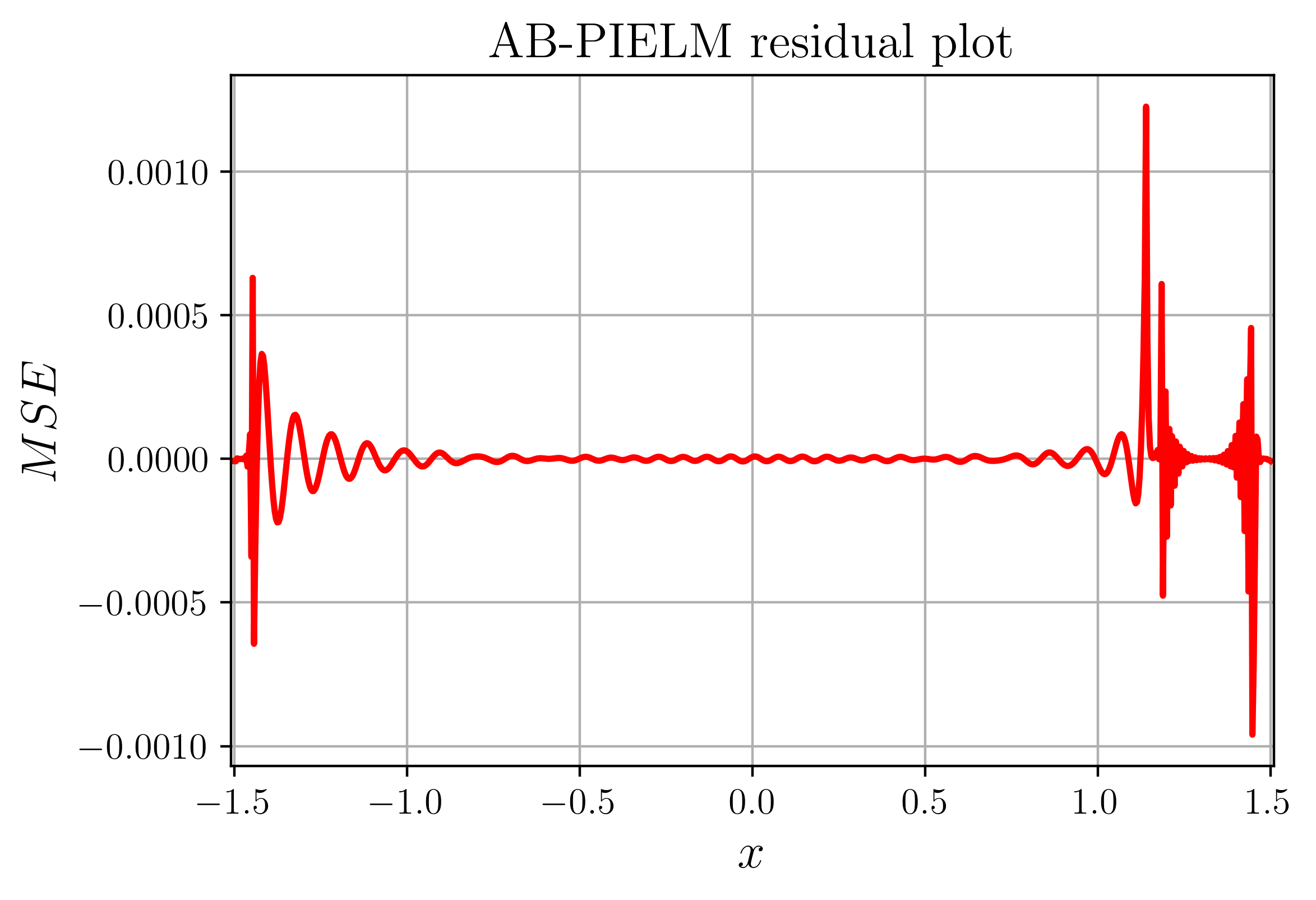}%
    \caption{}
    \label{fig:sinc_5domain_error}
  \end{subfigure}
  \caption{AB-PIELM approximation of the \emph{sinc} function using 5 subdomains (a) and the corresponding MSE residual plot (b).}
  \label{fig:sinc_5domain_combined}  
\end{figure}

Figure \ref{fig:sinc_elbow} shows the MSE residual as a function of the number of subdomains. An elbow is observed at 5 subdomains, 
which is selected as the optimal number. 
The AB-PIELM approximation of the \emph{sinc} function using 5 subdomains is shown in Figure \ref{fig:sinc_5domain}, while the MSE residual plot 
is shown in Figure \ref{fig:sinc_5domain_error}. 
Overall, the results demonstrate that the proposed framework can effectively approximate highly oscillatory functions by optimising 
the domain decomposition.

\begin{table}[h]
    \centering
    \caption{Comparison of performance metrics with existing literature for \emph{sinc} function approximation}
    \begin{tabular}{c|c|c}
          Metric & PINN-LM \citep{taylor_optimizing_2022} & AB-PIELM \\
        \hline
        Number of neurons &  9620 & 250 \\
        MSE & \num{1.64e-7} & \num{9.91e-9}\\
        \hline
    \end{tabular}
    
    \label{tab:placeholder}
\end{table}

\subsection{Problem 2: Advection–Diffusion}



\begin{figure}[h]
  \centering
  \begin{subfigure}{0.6\linewidth}
    \centering
    \includegraphics[width=\linewidth]{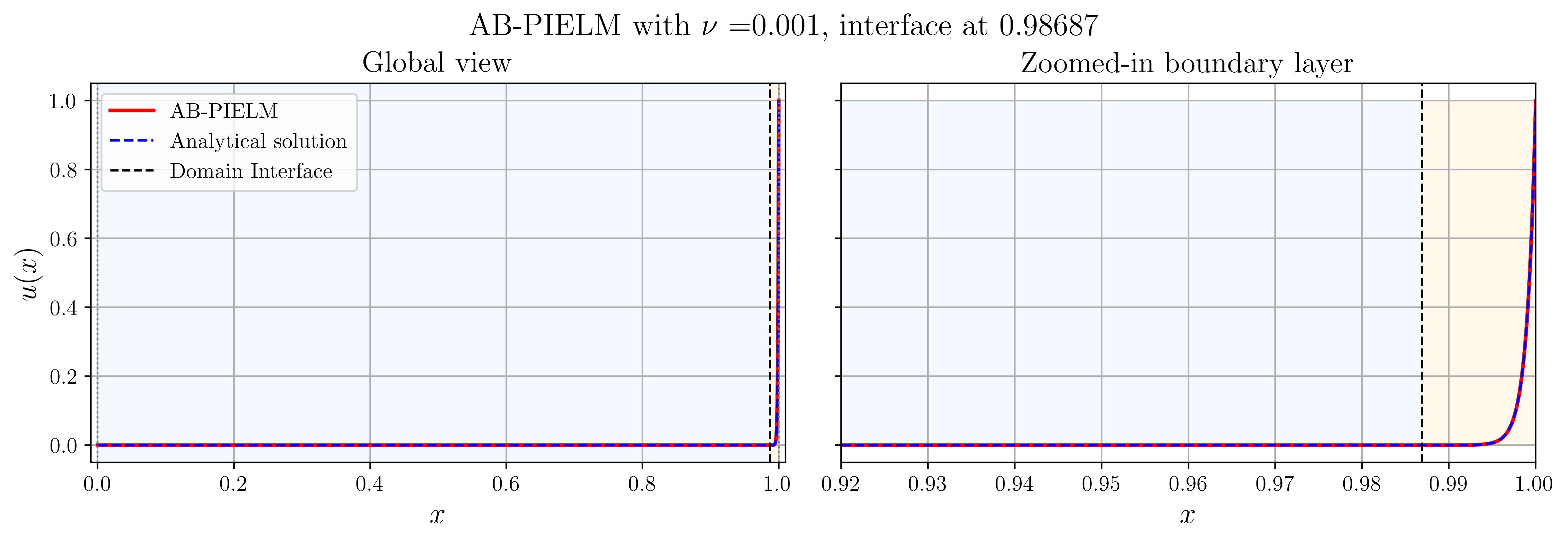}%
    \caption{}
    \label{fig:adv_diff_solution1}
  \end{subfigure}
  \begin{subfigure}{0.6\linewidth}
    \centering
    \includegraphics[width=\linewidth]{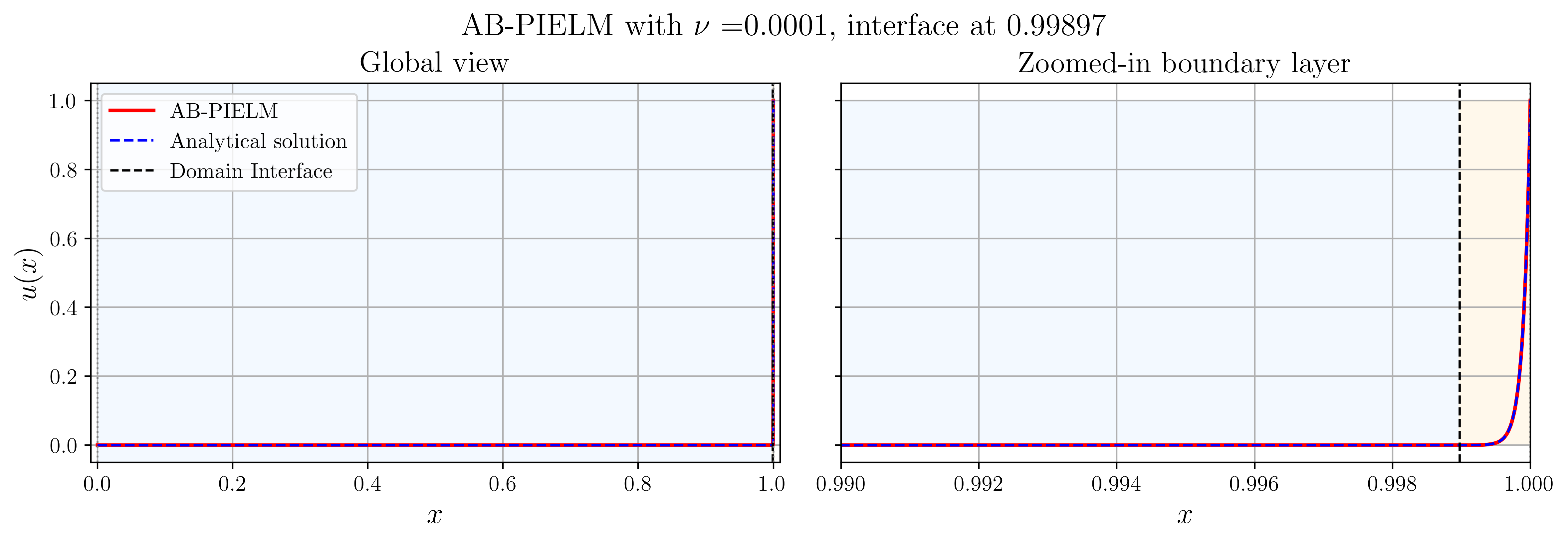}%
    \caption{}
    \label{fig:adv_diff_solution2}
  \end{subfigure}
  \begin{subfigure}{0.6\linewidth}

    \centering
    \includegraphics[width=\linewidth]{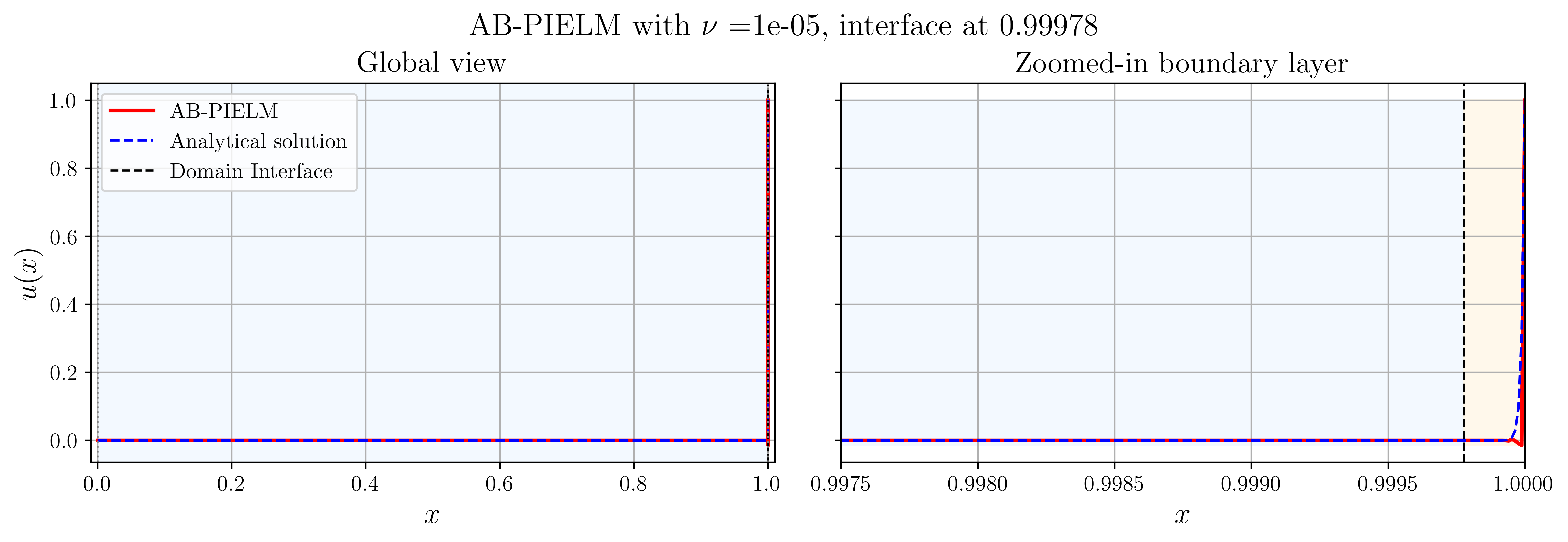}%
    \caption{}
    \label{fig:adv_diff_solution3}
  \end{subfigure}

  \caption{Advection–diffusion equation solutions for 
  different diffusion coefficients: 
  (a) $\nu = 10^{-3}$, (b) $\nu = 10^{-4}$, and (c) $\nu = 10^{-5}$. 
  The AB-PIELM solutions  are compared against the analytical solutions.}
  \label{fig:adv_diff_solutions}
\end{figure}

\begin{figure}[h]
  \centering
  \begin{subfigure}{0.6\linewidth}
    \centering
    \includegraphics[width=\linewidth]{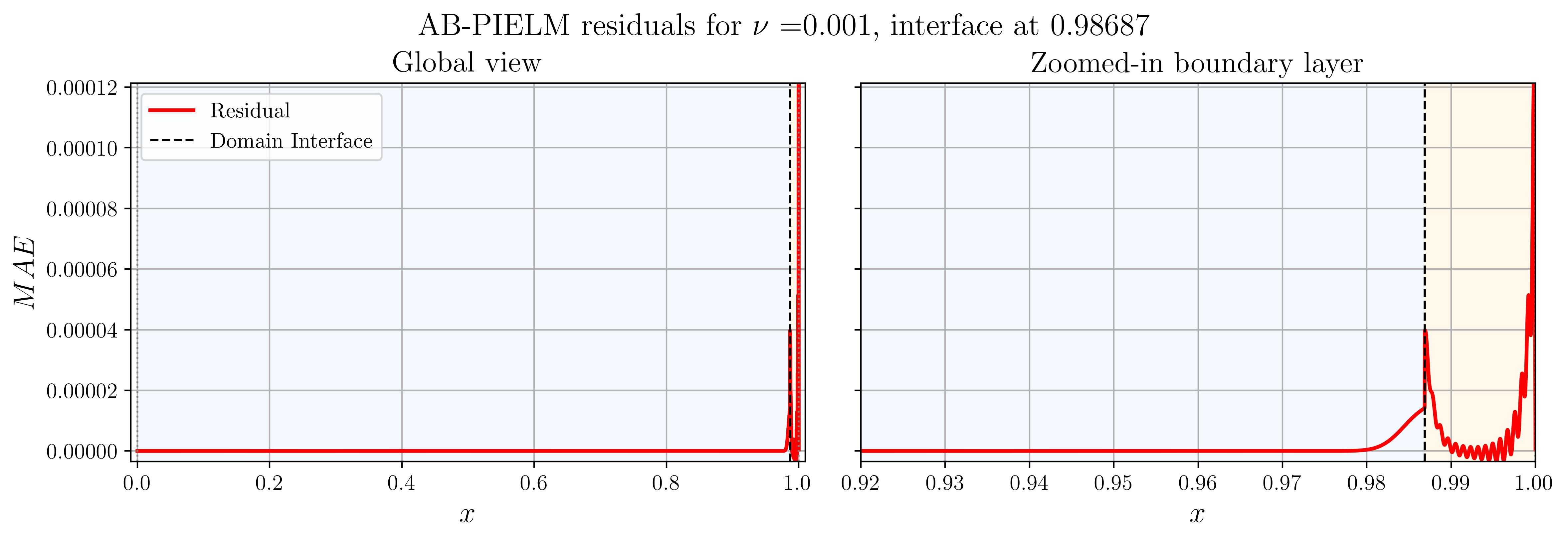}%
    \caption{}
    \label{fig:adv_diff_error1}
  \end{subfigure}

  \begin{subfigure}{0.6\linewidth}
    \centering
    \includegraphics[width=\linewidth]{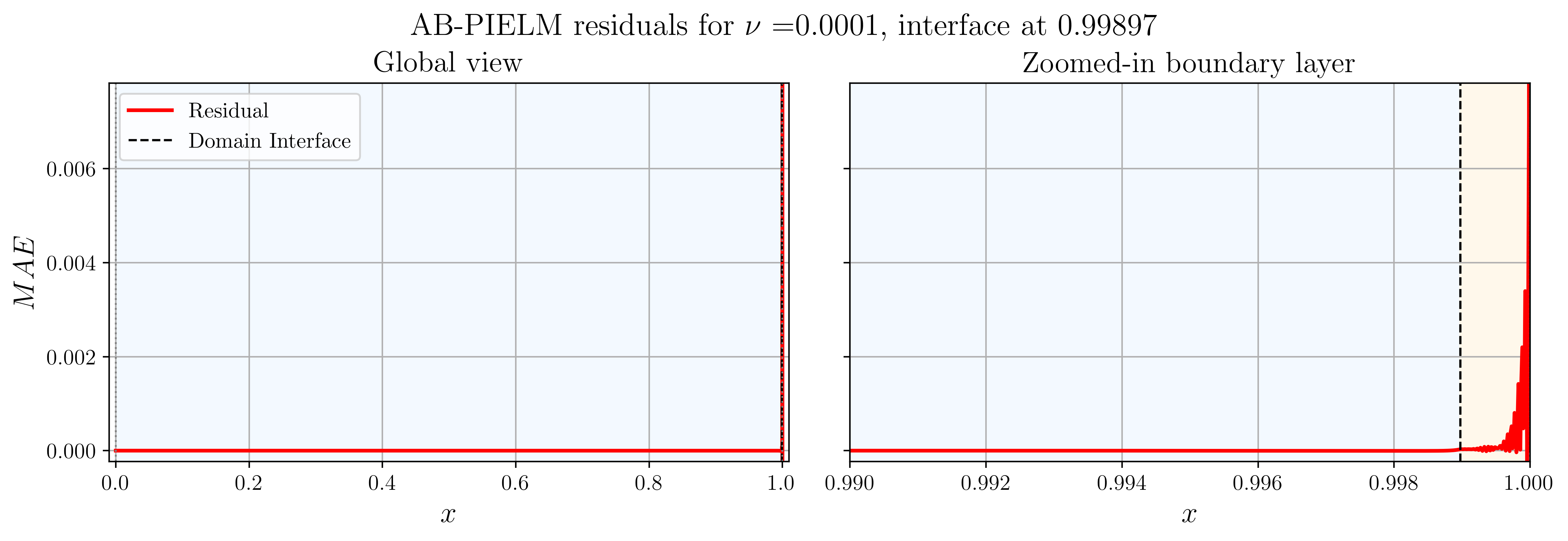}%
    \caption{}
    \label{fig:adv_diff_error2}
  \end{subfigure}

  \begin{subfigure}{0.6\linewidth}
    \centering
    \includegraphics[width=\linewidth]{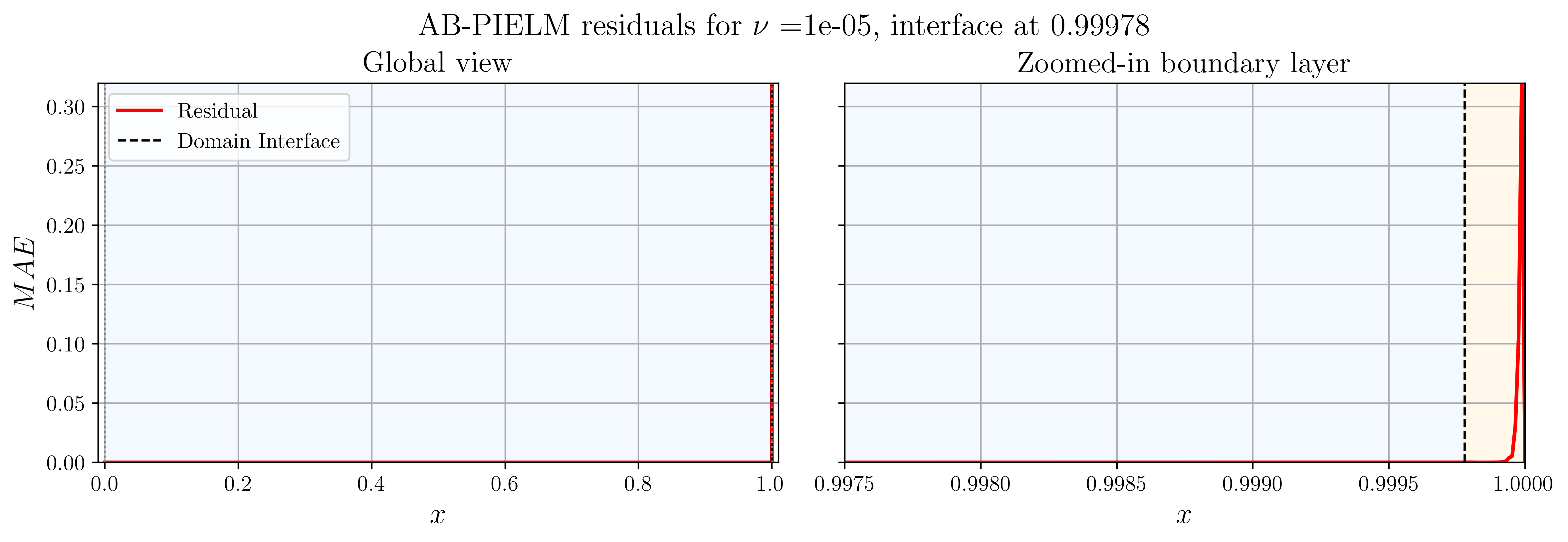}%
    \caption{}
    \label{fig:adv_diff_error3}
  \end{subfigure}

  \caption{MAE residual plots for the advection–diffusion equation for 
  different diffusion coefficients:
  (a) $\nu = 10^{-3}$, (b) $\nu = 10^{-4}$, and (c) $\nu = 10^{-5}$.}
  \label{fig:adv_diff_errors} 
\end{figure}

Figure \ref{fig:adv_diff_solutions} shows that the AB-PIELM solutions closely match the analytical solution of 
the advection–diffusion equation. The Mean Absolute Error (MAE) residual plot is shown in Figure \ref{fig:adv_diff_errors}. 
The performance of the AB-PIELM is also compared with other methods, namely PIELM and XTFC. These methods were the best performers
in a comparison of Neural Networks tested on the advection–diffusion problem
in \citep{de_florio_physics-informed_2024}. The performance of the vanilla distributed NEPIELM (DNEPIELM) is also reported for comparison.
The results are summarized in Table \ref{tab:adv_diff_summary}.

In summary, the AB-PIELM is significantly more accurate than the PIELM in all the values of $\nu$ considered, 
while using two orders of magnitude fewer neurons (PIELM is reported to 
entirely fail for the $\nu=0.0001$ case). The AB-PIELM also outperforms the XTFC method in the two
cases with the sharpest gradients ($\nu=0.0001$ and $\nu=0.00001$) while using two orders of magnitude fewer neurons, thereby demonstrating 
the better or comparable accuracy of the proposed method with significantly reduced model size.

\begin{table}[h]
  \centering
  \caption{Comparison of performance metrics number of neurons $L_{tot}$ and Mean Absolute Error for advection–diffusion equation at 
  different diffusion coefficient $\nu$ values using different methods from literature\citep{de_florio_physics-informed_2024}XTFC \citep{de_florio_physics-informed_2024}}
  \begin{tabularx}{0.9\textwidth}{Y|Y|c|c|c|c}
      $\nu$ &  Metric & PIELM  & XTFC & DNEPIELM & AB-PIELM \\
      \hline
      \multirow{2}{*}{$0.001$} & $L_{tot}$ & 10000 & 10000 & 100 & 100 \\
        & MAE & \num{4.89e-5} & \num{7.09e-13} & \num{2.25e-4} & \num{1.85e-7} \\
      \hline
      \multirow{2}{*}{$0.0001$} & $L_{tot}$ & 10000 & 10000 & 400 & 100 \\
        & MAE & \num{5.0e-1} & \num{1.39e-3} & \num{1.99e-4} & \num{4.56e-7} \\  
      \hline  
      \multirow{2}{*}{$0.00001$} & $L_{tot}$ & - & - & 400 & 100 \\
        & MAE & - & - & \num{4.37e-4} & \num{6.41e-6} \\
      \hline
  \end{tabularx}
        
  \label{tab:adv_diff_summary}
\end{table}

\subsubsection{Interpretibility of optimised hyperparameters}

The optimisation of hyperparameters in AB-PIELM reveals important insights into the model's behavior. 
The number of subdomains and the placement of interfaces are critical factors that influence the model's 
representation capacity and accuracy. Table \ref{tab:adv_diff_hyperparams} shows that for all three 
values of the diffusion coefficient $\nu$, the optimised interface is located close to the sharp gradient region
thus maximising the representation capacity where it is most needed. It is also observed
that the $\sigma$ values in the left domain (smooth region) are significantly larger than those in the right domain (sharp gradient region).
This indicates that the model requires a finer resolution (smaller $\sigma$) to accurately capture the 
steep gradient, while a coarser resolution (larger $\sigma$) suffices in the smooth region.




\begin{table}[h]
  \centering
  \caption{Comparison of hyperparameters over domain 1 (left) containing the smooth region and domain 2 (right) containing the sharp gradient for advection–diffusion equation at different diffusion coefficients}
  \begin{tabular}{c|c|c|c}
      Diffusion coefficient & Domain & Domain length & $\sigma$\\
      \hline
      \multirow{2}{*}{$\nu=0.001$} & Domain 1 & 0.9869 & 0.0025 \\
        & Domain 2 & 0.0131 & \num{9.311e-4} \\ 
      \hline
      \multirow{2}{*}{$\nu=0.0001$} & Domain 1 & 0.9990 & 0.0026 \\
        & Domain 2 & 0.0010 & \num{4.310e-5} \\
      \hline
      \multirow{2}{*}{$\nu=0.00001$} & Domain 1 & 0.99999 & 0.1715 \\
        & Domain 2 & 0.00001 & \num{5.509e-8} \\
      \hline
  \end{tabular}
  \label{tab:adv_diff_hyperparams}
\end{table}

\subsection{Inverse Problem Results}

Table \ref{tab:inverse_results} shows the recovered diffusion coefficient $\nu^*$ for different noise levels in the synthetic data. 
AB-PIELM is able to accurately recover the true diffusion coefficient  even in the presence of noise upto 5\%, demonstrating the robustness of the method for inverse problems. The results also show that as the noise level increases, the recovered $\nu^*$ deviates more from the true value, which is expected due to the increased uncertainty in the data. However, even at a noise level of $\sigma^2_{noise} = 0.05$, the recovered $\nu^*$ remains reasonably close to $\nu_{true}$, indicating that AB-PIELM can effectively handle noisy data in inverse problem settings.

\begin{table}[h]
  \centering
  \caption{Recovered diffusion coefficient $\nu$ for different noise levels}
  \begin{tabular}{c|c|c}
      $\nu_{true}$ & $\sigma^2_{noise}$ & Recovered $\nu^*$ \\
      \hline
      \multirow{3}{*}{$0.005$} & $0$ & 0.0049172 \\
        & $0.01$ & 0.0049917 \\
        & $0.05$ & 0.0054590 \\
      \hline
      \multirow{3}{*}{$0.0005$} & $0$ & 0.0005181 \\
        & $0.01$ & 0.0005236 \\
        & $0.05$ & 0.0004491 \\
      \hline
  \end{tabular}
  \label{tab:inverse_results}
\end{table}

\paragraph{Limitations}
\begin{itemize}
\item \textbf{Linear operators:} The normal–equation formulation relies on linearity of the PDE in the solution coefficients. Nonlinear PDEs would require outer iterative linearization, reducing the deterministic training advantage.

\item \textbf{Scaling to high dimensions:} Domain decomposition introduces interface constraints whose number grows rapidly with dimension, leading to larger coupled systems and increased memory cost.

\item \textbf{Unsteady problems:} The present formulation addresses steady equations. Extending to time-dependent PDEs would require time marching or space–time decompositions, significantly increasing system size and conditioning complexity.

\item \textbf{Penalty parameters:} The weights in the physics-informed objective must be specified a priori since they define the objective itself and cannot be optimized within the same loss.
\end{itemize}

\section{Conclusion}

This work introduced a distributed adaptive physics-informed extreme learning machine that augments the normal-equation PIELM formulation 
with optimisable spatial and spectral hyperparameters. By allowing both the domain partition and the locality of the basis functions to 
adapt to the solution structure, the method addresses a central limitation of global neural PDE solvers: the inability to efficiently 
represent localized behaviour using a uniform approximation space. The resulting formulation retains deterministic training while 
enabling problem-dependent allocation of representational capacity. Numerical experiments demonstrated accurate approximation of 
oscillatory targets and robust solutions of singularly perturbed advection–diffusion equations, as well as stable recovery of the 
diffusion coefficient from sparse and noisy observations. The learned hyperparameters were also interpretable, with interfaces 
concentrating near sharp gradients and basis widths varying according to local solution smoothness.





\bibliography{refs_arxiv}

\bibliographystyle{iclr2026_conference}

\end{document}

%% file: math_commands.tex
\usepackage{amsmath,amsfonts,bm}

\def\eqref#1{equation~\ref{#1}}

\def\1{\bm{1}}

\DeclareMathAlphabet{\mathsfit}{\encodingdefault}{\sfdefault}{m}{sl}
\SetMathAlphabet{\mathsfit}{bold}{\encodingdefault}{\sfdefault}{bx}{n}

\DeclareMathOperator*{\argmin}{arg\,min}